\documentclass[aps,pre,twocolumn,showpacs,preprintnumbers,amsfonts,amssymb,superscriptaddress]{revtex4}
\usepackage{placeins}
\usepackage{graphicx}
\usepackage{times}
\usepackage{amssymb}
\usepackage{amsmath}
\usepackage{amsfonts}
\usepackage{color}
\begin{document}
\title{Coevolutionary dynamics in large, but finite populations}
\author{Arne Traulsen}
\affiliation{Program for Evolutionary Dynamics,
Harvard University,
One Brattle Square,
Cambridge, MA 02138, USA} 
\author{Jens Christian Claussen}
 \affiliation{Institut f{\"u}r Theoretische Physik und
Astrophysik, Christian-Albrechts Universit{\"a}t,
Olshausenstra{\ss}e 40, 24098 Kiel, Germany}
\author{Christoph Hauert}
\affiliation{Program for Evolutionary Dynamics,
Harvard University,
One Brattle Square,
Cambridge, MA 02138, USA} 
\preprint{Published in: ~ Physical Review E {\bf 74}, 011901 (2006)}
\date{12 April 2006}
\pacs{87.23-n,89.65.-s,87.23.Kg
DOI: 11.1103/PhysRevE.74.011901 
}

\begin{abstract}
Coevolving and competing species or game-theoretic strategies exhibit rich and complex 
dynamics for which a general theoretical framework based on finite populations is still lacking. 
Recently, an explicit mean-field description in the form of a Fokker-Planck equation was derived  
for frequency-dependent selection with two strategies in finite populations based on microscopic processes 
[A.~Traulsen, J.~C. Claussen, and C.~Hauert, Phys.\ Rev.\ Lett.\ {\bf 95}, 238701 (2005)].
Here we generalize this approach in a twofold way: First, we extend the framework to an 
arbitrary number of strategies and second, we allow for mutations in the evolutionary process. The 
deterministic limit of infinite population size of the frequency dependent Moran process yields 
the adjusted replicator-mutator equation, which describes the combined effect of selection and mutation. 
For finite populations, we provide an extension taking random drift into account.  
In the limit of neutral selection, i.e.\ whenever the process is determined by random drift and mutations, 
the stationary strategy distribution is derived. This distribution forms the background for the 
coevolutionary process. In particular, a critical mutation rate $u_c$ is obtained separating two scenarios: 
above $u_c$ the population predominantly consists of a mixture of strategies whereas below $u_c$ the 
population tends to be in homogenous states. For one of the fundamental problems in evolutionary biology, 
the evolution of cooperation under Darwinian selection, we demonstrate that the analytical framework 
provides excellent approximations to individual based simulations even for rather small population sizes. 
This approach complements simulation results and provides a deeper, systematic understanding of 
coevolutionary dynamics.
\vspace*{-2ex}
\end{abstract}
\maketitle
\vspace*{-6ex}
\section{Introduction}
\vspace*{-1ex}
Darwinian evolution incorporates three basic processes: mutation, selection and random drift. In evolving populations mutations create variation, which produce fitness differences for selection to act upon and random drift accounts for stochastic effects arising due to the finite size of the population. 
The consideration of finite populations has a long tradition in population genetics 
\cite{fisher:1930,wright:1931,moran:1962ef, kimura:1968aa} with frequency independent selection. Often, finite size effects can be viewed as corrections to a continuum theory based on an infinite population \cite{peng:2004aa,cohen:2005aa}. Here, we analyze such finite population effects for frequency dependent selection. Under frequency dependent selection, or co-evolutionary dynamics, the fitness of each individual is affected by interactions with other members of the population. The corresponding mathematical framework is evolutionary game theory \cite{maynard-smith:1982to,nowak:2004aa}. In the absence of mutations and in the deterministic limit of large well-mixed populations where individuals interact randomly, the replicator equation \cite{taylor:1978wv,hofbauer:1998mm} describes the change in frequency of the different strategic types in the population. Including mutations in this limit leads to the replicator-mutator equation \cite{bomze:1995rm,page:2002an}. Interestingly, results can be quite different when turning to coevolutionary processes in finite populations based on the Moran process \cite{moran:1962ef,nowak:2004pw,taylor:2004wv}. The frequency dependent Moran process is a stochastic birth-death process where an individual is randomly selected for reproduction with a probability proportional to its fitness. The clonal offspring then replaces a randomly chosen individual in the population such that the population size remains constant. If there are only two types of individuals in the population, the evolutionary process can be mapped onto a random walk in one dimension \cite{antal:2005aa}. For more than two types the dynamics becomes significantly more difficult \cite{imhof:2005oz}. Further complications arise when individuals no longer interact randomly but, instead, spatial structure leads to limited local interactions. Traditionally, this is modeled by arranging individuals on a spatial lattice \cite{nowak:1992pw,lindgren:1994to,szabo:1998wv} or on more general networks \cite{ebel:2002aa,szabo:2002mf,vukov:2005fa,santos:2005bb,santos:2006pn,ohtsuki:2006na} and leads to new phenomena, which includes critical phase transitions \cite{szabo:2002te}, and can affect the evolutionary process even in frequency independent settings \cite{lieberman:2005qx,antal:2006pr}.
\\ \indent
For two strategic types in well-mixed populations we established an explicit connection between microscopic stochastic processes in finite populations and the deterministic limit of infinite populations \cite{traulsen:2005hp}. The finite size effects are captured in the drift and diffusion terms of a Fokker-Planck equation where the diffusion term vanishes with $1/N$ for increasing population sizes. In the present paper the Fokker-Planck equation is generalized to an arbitrary number of strategies as well as to cover mutational changes of the strategic type. 
\\ \indent
In Sec.\ \ref{gmps}, the frequency dependent Moran process is generalized to $d$ strategies including mutations. The derivation of the corresponding Fokker-Planck equation is recalled in Sec.\ \ref{fpes} and the stationary distribution is derived in Sec.\ \ref{sds}. This covers frequency independent and frequency dependent selection. As applications, we consider the Prisoner's Dilemma and the Snowdrift game. The relation to dynamics in infinite populations is derived in Sec.\ \ref{repmuts}. An example for an alternative microscopic process is provided in Sec.\ \ref{lurs}.
\vspace*{-2ex}
\section{Generalized Moran process}
\label{gmps}
\vspace*{-1ex}
The frequency dependent Moran process has been introduced as a stochastic process among two types  \cite{nowak:2004pw}. Although the extension to more strategies is straightforward \cite{imhof:2005oz},
the dynamics becomes significantly more difficult. 
 In complete analogy to the Moran process among two individuals, an individual is chosen at random proportional to fitness in every time step. This individual produces offspring which replaces a randomly chosen individual. In addition to the original frequency dependent Moran process, we include mutations, i.e.\ an individual of type $k$ produces offspring of type $j$ with probabiliy $q_{jk}$, where $\sum_{j=1}^d q_{jk}=1$ and $d$ is the number of different types in the population. The special case of vanishing mutations is recovered by setting $q_{jk}= \delta_{jk}$. 
Note that there are no restrictions on the frequency of mutations or on the states in which mutations occur. For $d>2$, the state space of the system is not longer a simple one-dimensional chain, but the discretized simplex $S^N_d=\lbrace(i_1,\ldots,i_d) \in N_0^d: \sum_{j=1}^d i_j = N\rbrace$, where $i_j$ is the number of individuals of type $j$ and $N$ is the total size of the population.
The fitness $\pi_j$ of an individual of type $j$ is a linear combination of the payoff from interactions given by the entries of the $d \times d$ payoff matrix $M$ with elements $m_{jk}$ and a baseline fitness, which is set to one for convenience. Hence,  
\begin{equation}
\pi_j(i_1,\ldots,i_d) = 1-w+w\frac{\sum_{k=1}^d m_{jk} i_k-m_{jj}}{N-1}, 
\end{equation}
where self interactions are excluded by subtracting the payoff $m_{jj}$.  The selection intensity $w$ determines the relative contributions of the baseline fitness and the interactions to the total fitness of an individual. For $w \to 0$ selection is weak and payoffs from the game represent small perturbations to the constant background fitness. For strong selection, $w \to 1$, the influence of the background fitness vanishes and fitness is determined entirely by interactions. 
Since individuals are selected {\em proportional} to fitness, the fitness has to be a positive number. This can be guaranteed by payoff matrices with positive entries only, $m_{jk} \geq 0$ for all $j$ and $k$, or by a sufficiently small $w$.

Based on this fitness definition, the probability $T_{k  j}$ that an individual of type $k$ is replaced by an individual of type $j$ can be calculated. $T_{k  j}$ is the product of the probability that the individual chosen at random for elimination is of type $k$ and the probability that an offspring of type $j$ is produced. There are two possibilities to produce such an offspring:  An individual of type $j$ is chosen for reproduction with probability $i_j \pi_j/ (N \phi$), where
$\phi = \sum_{m=1}^d  \pi_m i_m/N$,
is the average payoff in the population. 
It produces identical offspring (with probability $q_{jj}$).
The second possibility is that an individual of type $l$ produces an offspring of type $j$ due to mutations (with probability $q_{lj}$). Hence, the probability $T_{k  j}$ is given by
\begin{equation}
T_{k  j}(i_1,...,i_d) = 
 \frac{1}{
 N \phi } \sum_{l=1}^d i_l \pi_l(i_1,\ldots,i_d) \; q_{lj}  \frac{i_k}{N}.
 \label{Morantransprob}
\end{equation}
These transition probabilities describe the effect of selection and mutation in finite populations. The probabilistic nature of this coevolutionary process accounts for random drift. 

Since the state space of this system is no longer a one-dimensional chain for $d>2$, many standard methods can no longer be applied. However, as in the case $d=2$ a Fokker-Planck equation for the dynamics of the system can be derived for large populations.

Although we are motivated by the generalized frequency dependent Moran process, the calculations following in Secs.\ \ref{fpes} and \ref{sds} are valid in a more general sense. In particular, they apply to any coevolutionary birth death process, such as the local update rule discussed in \cite{traulsen:2005hp}, see Sec.\ \ref{lurs}.

\section{Fokker-Planck equation}
\label{fpes}
\vspace*{-1ex}
In analogy to Ref.\  \cite{traulsen:2005hp} for $d=2$, a general Fokker-Planck equation for 
the probability density of strategies $\rho({\boldsymbol x})$ in an evolutionary game among $d$ types can be derived. The constant population size of the generalized Moran process leads to a normalization of ${\boldsymbol x}$, $\sum_{j=1}^d x_j =1$. Thus, we have $d-1$ independent variables. As shown in the Appendix, a Kramers-Moyal expansion yields the Fokker-Planck equation:
\begin{equation}
\label{FPE}
\dot \rho({\boldsymbol x}) =
- \sum_{k=1}^{d-1} \frac{\partial}{\partial x_k} \rho({\boldsymbol x}) a_{k}({\boldsymbol x})
+\frac{1}{2} \! 
\sum_{j,k=1}^{d-1} \frac{\partial^2}{\partial x_k \partial x_j}\rho({\boldsymbol x}) 
b_{jk}({\boldsymbol x}),
\end{equation}
where ${\boldsymbol x} = (x_1,\ldots, x_d)$ with $x_j = i_j/N$.  
The drift coefficients are given by
\begin{equation}
a_{k}({\boldsymbol x}) = \sum_{j=1}^d T_{j k}({\boldsymbol x}) - T_{k j}({\boldsymbol x})
\label{drift}
\end{equation}
and can be interpreted as an effective flow into state $k$: The first term is the probability
that the number of $k$ individuals increases, whereas the second term describes the 
probability for transitions from $k$ to other states. For the diffusion matrix, we find
\begin{equation}
b_{jk}({\boldsymbol x}) \!\!
= \!\! \frac{1}{N} \!\!\left[ -T_{j k}({\boldsymbol x}) - T_{k j}({\boldsymbol x}) +\delta_{jk} \sum_{l=1}^d
T_{j l}({\boldsymbol x})+T_{l j}({\boldsymbol x}) \!
 \right].
 \end{equation}
For two types, $d=2$, and without mutations, the Fokker-Planck equation from \cite{traulsen:2005hp} is recovered. 
Eq.\ (\ref{FPE}) describes the dynamics on the basis of changes in the probability distribution. Traditionally, coevolutionary systems are often described by considering dynamical equations for the state of the system. General conclusions are then based on averages made over several realizations of the process.  Here, the correspondence between both descriptions becomes clear, since the Fokker-Planck equation (\ref{FPE}) corresponds to a stochastic differential equation \cite{gardiner:1985bv,honerkamp:1994mm,kampen:1997xg}. The noise arises only from stochastic updating and is therefore not correlated in time. Hence, the It{\^o} calculus can be applied to derive 
a Langevin equation describing the development of $x$. Equation (\ref{FPE}) corresponds to the stochastic replicator-mutator equation
\begin{equation}
\dot x_k = a_k({\boldsymbol x}) + \sum_{j=1}^{d-1} {c_{kj}({\boldsymbol x})} \xi_j(t) 
\label{langevin}
\end{equation}
where $ {c_{kj}}$ is defined by $\sum_{l=1}^{d-1} {c_{kl}({\boldsymbol x})} {c_{lj}({\boldsymbol x})} =  b_{kj}({\boldsymbol x})$.
Each element of the vector ${\boldsymbol \xi}$ is Gaussian white noise with unit variance. The different elements are uncorrelated, $\langle \xi_k(t) \xi_j(s) \rangle = \delta_{kj} \delta(t-s)$. Equation (\ref{langevin}) allows to approximate fluctuations arising from finite populations by Langevin terms that appear in a replicator equation. For any given payoff matrix, selection pressure, population size and mutation rate, it provides a quantitative description of the fluctuations introduced by stochastic microscopic update processes. 

\section{Stationary distribution}
\label{sds}
\vspace*{-1ex}
The stationary distribution of strategies $\rho^{\ast}({\boldsymbol x})$ can be derived from the 
Fokker-Planck equation (\ref{FPE}), which can be written in the form
\begin{equation}
\label{flowequation}
\dot \rho({\boldsymbol x}) = 
-\nabla {\boldsymbol J}({\boldsymbol x}), 
\end{equation}
where the $d-1$ elements of the probability current  ${\boldsymbol J}({\boldsymbol x})$ are 
\begin{equation}
J_k({\boldsymbol x}) =  \rho({\boldsymbol x}) a_{k}({\boldsymbol x})
- \frac{1}{2}  
\sum_{j=1}^{d-1} \frac{\partial}{\partial x_j}\rho({\boldsymbol x}) 
b_{jk}({\boldsymbol x}).
\end{equation}
In general, the stationary solution $\dot \rho({\boldsymbol x})= 0$ is equivalent to a probability current without sources. A special case of this is a probability current vanishing everywhere, i.e.\ $J_k({\boldsymbol x})=0$ for all ${\boldsymbol x}$. This leads to
\begin{equation}
\ \sum_{k=1}^{d-1} b_{jk}({\boldsymbol x}) \frac{\partial }{\partial x_k} \rho^{\ast}({\boldsymbol x})
= \rho^{\ast}({\boldsymbol x}) \! \left[2 a_j({\boldsymbol x}) \!-\! \sum_{k=1}^{d-1}  \frac{\partial}{\partial x_k}  b_{jk}({\boldsymbol x})\right].
\label{rhoeq}
\end{equation}
If we exclude the degenerate cases with $\det b_{jk} ({\boldsymbol x})= 0$, the matrix $b_{jk}({\boldsymbol x})$ can be inverted and Eq.~(\ref{rhoeq}) reduces to 
\begin{equation}
\frac{\partial}{\partial x_i} \! \ln  \rho^{\ast}({\boldsymbol x}) \!\!=\!\! \sum_{j=1}^{d-1} b^{-1}_{ij}({\boldsymbol x}) 
\!\left[2 a_j({\boldsymbol x}) \! - \! \sum_{k=1}^{d-1}  \frac{\partial}{\partial x_k}  b_{jk}({\boldsymbol x})\right] 
\!\!=\! \Gamma_i({\boldsymbol x}) 
\end{equation}
A solution of this equation only exists if ${\boldsymbol \Gamma} ({\boldsymbol x})$ is a gradient \cite{kampen:1997xg,gardiner:1985bv}. 
In that case, the stationary solution of Eq.\ (\ref{FPE}) can be obtained from a line integral between ${\boldsymbol x_0}$ and ${\boldsymbol x}$ (which is independent of the path) 
\begin{equation}
\rho^{\ast}({\boldsymbol x}) ={\cal N} \exp\left[ 
\int_{\boldsymbol x_0}^{\boldsymbol x} {\boldsymbol \Gamma} 
\left({\boldsymbol y} \right) \cdot d{\boldsymbol y}
\right],
\label{stathighd}
\end{equation}
where $\cal N$ is a normalization constant. However, in co-evolutionary systems the requirement that ${\boldsymbol \Gamma}({\boldsymbol x})$ is a gradient is often not fulfilled.
In that case, a stationary solution may still exist in which the ansatz of a vanishing probability current, ${\boldsymbol J}({\boldsymbol x}) ={\boldsymbol 0}$, is not valid although the divergence of ${\boldsymbol J}({\boldsymbol x})$ remains zero, $\nabla {\boldsymbol J}({\boldsymbol x})=0$. In these cases the stationary distribution has to be derived by different means. 

A particularly simple case where these problems do not arise is $d=2$, where only two types $A$ and $B$ are present. The state 
of the system can be described by the fraction $x$ of $A$ individuals. The remaining fraction $1-x$
consists of $B$ individuals. Assuming a mutation rate $u$ from $A$ to $B$ as well as from $B$ to $A$, 
the transition probabilities read 
\begin{eqnarray}
 T_{BA}({x})  & = &  \frac{ x \pi_A(x) (1-u)+(1-x) \pi_B(x) u}{\phi} (1-x),\nonumber \\
 T_{AB}({x})  & = &   \frac{ x \pi_A(x) u + (1-x) \pi_B(x) (1-u)}{\phi} x.
 \end{eqnarray}
The drift and diffusion terms in the Fokker-Planck equation become
\begin{eqnarray}
a(x) & = &  x (1-x)  \frac{\pi_A(x)-\pi_B(x)}{\phi} \\ \nonumber 
&& + u  \frac{ (1-x) \pi_B(x)- x \pi_A(x)}{\phi} \\
b(x) & = &  \frac{1}{N} x (1-x) \frac{\pi_A(x)+\pi_B(x)}{\phi}  \\ \nonumber 
&& + \frac{u}{N} \frac{ x \pi_A(x) +(x-1) \pi_B(x)}{\phi}(2x-1). 
\end{eqnarray}
The stationary solution of Eq.\ (\ref{FPE}) can now be computed as outlined above and is given by
\begin{equation}
\label{statdist}
\rho^{\ast}(x) = {\cal N} \exp \left[\int_0^x \Gamma(y) dy \right],
\end{equation}
where
$\Gamma(y)=b^{-1}(y)(2 a(y)- b'(y))$,
$b'(y) = \frac{d}{dy} b(y)$ and ${\cal N}= \int_0^1 \exp \left[\int_0^x \Gamma(y) dy \right] dx$. 
In general, the derivation of a stationary distribution of strategies under the effect of mutations cannot be solved analytically. Earlier approaches have approximated such distributions either by assuming weak mutations \cite{imhof:2006ee}, where only transition probabilities between absorbing states have to be considered, or by neglecting all mutations except those in the absorbing states \cite{claussen:2005eh}. The distribution given by Eq.\ (\ref{statdist}) for large $N$ is valid for arbitrary selection intensity $w$, mutation rate $u$, and any payoff matrix. 

\subsection{Neutral dynamics}
\vspace*{-1ex}
In the frequency dependent Moran process, often weak selection, $w \ll 1$, is considered \cite{nowak:2004pw,ohtsuki:2006na}. For weak selection, the interactions are only corrections to a background of neutral dynamics \cite{kimura:1968aa,crow:1970ck}.
Therefore, we first discuss the stationary distribution for $w=0$, where the game has no influence on the fitness.  In this case, ${\boldsymbol \Gamma}({\boldsymbol x})$ is a gradient, which is equivalent to the assumption that the probability current vanishes everywhere, and the stationary distribution can be computed directly from Eq.~(\ref{stathighd}). 
For $d=2$, we find for the drift term 
$a(x) = u(1-2x)$ and for the diffusion term
$b(x) = {(u(2x-1)^2+2x(1-x))}/{N}$. Both results are valid for the case in which the mutation rates from $A$ to $B$ and vice versa are $u$. Hence, $\Gamma(x)$ is given by
\begin{equation}
\Gamma(x)=-2 \frac{(u (2+N) -1) (2 x-1)}{u (2 x-1)^2+2x (1-x) }.
\end{equation}
\begin{figure}[thbp]
\includegraphics[totalheight=9cm,angle=270]{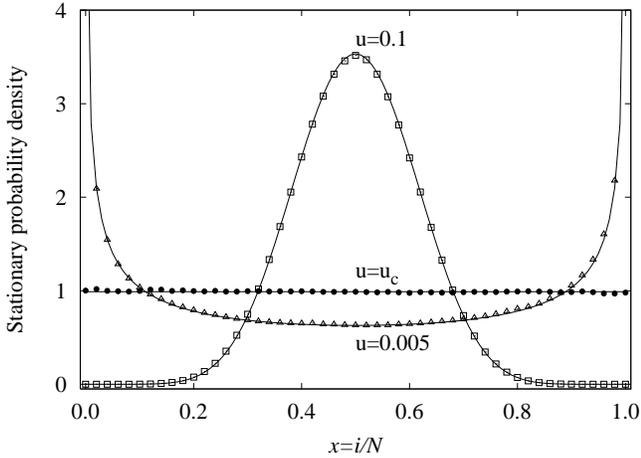}
\caption{
For neutral selection, $w=0$, three different scenarios occur for the stationary distribution of the system: 
For  $u<u_c=1/(2+N)$, the system spends most of the time near the absorbing states, leading to maxima of the stationary distribution at $i=0$ and $i=N$, as shown for $u=0.005$. 
For $u=u_c \approx 0.01$, the probability to leave the absorbing states due to mutations and the probability to reach them by random drift are approximately equal, leading to a uniform distribution. 
For $u>u_c$, there is on average more than one mutation per generation and the stationary distribution is centered around $i=N/2$. In all three cases, numerical simulations of the stationary distribution depicted by symbols agree very well with the theoretical result Eq.~(\ref{statdist}) shown as lines (simulations are averages over $10^8$ time steps, $N=100$).  
}
\label{neutral}
\end{figure}
$\Gamma(x)$ vanishes for the critical mutation rate $u_c = 1/(2+N)$. 
For $u<u_c$, $\Gamma(x)$ has a minimum at $x=1/2$. Accordingly, the stationary probability distribution has maxima at the boundaries when only few mutations per generation occur. On the other hand, for $u>u_c$ there is
a maximum at $x=1/2$. Mutations lead the system away from the states  $x=0$ and $x=1$ and the distribution is centered around the middle where both strategies have equal frequencies, see Fig. \ref{neutral}. For $u=u_c$ we have $\Gamma(x)=0$ and hence Eq.\ (\ref{statdist}) predicts a uniform distribution. However, simulations show a distribution that decays slightly toward the boundaries for small $N$ (cf.\ Fig.\ \ref{neutral}). Strictly speaking, a uniform distribution is observed only in the limit $N \to \infty$.

\begin{figure}[htbp]
\includegraphics[totalheight=4.2cm]{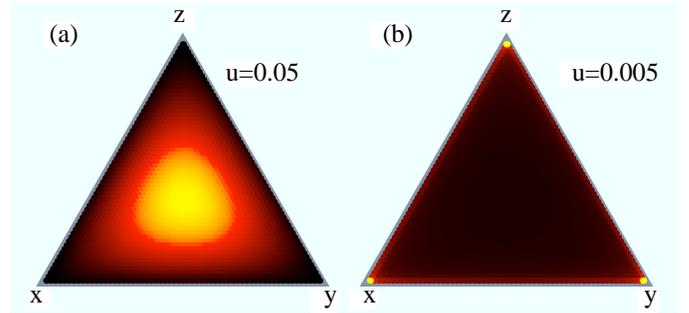}
\caption{
(Color online) 
For $d=3$ strategies and neutral selection ($w=0$), the stationary distribution in the strategy space spanned by the simplex $S_3$ is shown encoded by a color scale, where bright colors indicate high values. 
(a) For mutation rates higher than the critical mutation rate $u_c = 1/(3+N)$, mutations drive the system away from the absorbing states at the corners of the simplex. Due to the symmetry of the system, the stationary distribution has a maximum centered in the middle of the simplex where $x=y=z$. 
(b) If the mutation rates are smaller than than the critical mutation rate $u_c$, the system spends considerable time in the absorbing states at the corners. Occasionally, the edges are reached by mutations. However, since the system typically reaches the corners again before the next mutation occurs, the stationary probability density in the interior is very small. 
In both cases, the simulations of the stationary probability distribution shown here (averages over $10^8$ time steps) do not deviate significantly from the analytical result Eq.\ (\ref{stathighd}): for $u=0.05$ the maximal deviation is $\approx 2 \%$ and for $u=0.005$ it is $\approx 6\%$ (population size $N=60$). 
}
\label{neutrald}
\end{figure}

Critical mutation rates can also be computed for general $d$ in the case of symmetric mutations in which the mutation rate between two different states is $u$, i.e.\ $q_{lj} = u + \delta_{lj}(1-du)$. 
All elements of ${\boldsymbol \Gamma}({\boldsymbol x})$ vanish for 
$u=u_c=1/(d+N)$, because
$2 a_j({\boldsymbol x}) = \sum_{k=1}^{d-1}  \frac{\partial}{\partial x_k}  b_{jk}({\boldsymbol x})$. Hence, a uniform distribution is expected for $u=u_c$ in the limit of $N \to \infty$. For smaller mutation rate $u<u_c$, the distribution has maxima at the corners of the simplex where only one strategy is present. For larger $u>u_c$, the distribution has one maximum in the interior of the simplex. Fig.~\ref{neutrald} illustrates these two different cases for neutral selection with three different types, $d=3$.

\subsection{Frequency dependent selection}
\vspace*{-2ex}
For $w>0$, interactions lead to frequency dependent transition probabilities and to corrections of the stationary distribution. Depending on the nature of the game and the strength of selection, significant differences from neutral evolution are possible.

As examples, we consider two generic cases of $2 \times 2$ games, the Prisoner's Dilemma and the Snowdrift game. 
In the Prisoner's Dilemma \cite{rapoport:1965pd,axelrod:1984yo}, two players choose simultaneously whether to cooperate or to defect. 
The cost of cooperation is $c$, while the benefit from cooperation is $b$. The interaction of a cooperator with a defector leads to a payoff of $-c$ for the cooperator and $b$ for the defector. When two cooperators interact, they both get the payoff $b-c$, whereas the interaction of two defectors leads to a payoff of $0$. Thus, irrespective of the other player's move, it is always better to defect. However, the reward for mutual cooperation would be the mutually preferred outcome and hence the dilemma. The game is described by the payoff matrix
\begin{equation}
M = 
\left(
\begin{array}{cc} 
m_{11} & m_{12}  \\
m_{21} & m_{22} 
\end{array}
\right)
=
\left(
\begin{array}{cc} 
b-c & -c  \\
b & 0 
\end{array}
\right),
\end{equation}
\begin{figure}[thbp]
\includegraphics[totalheight=9cm,angle=270]{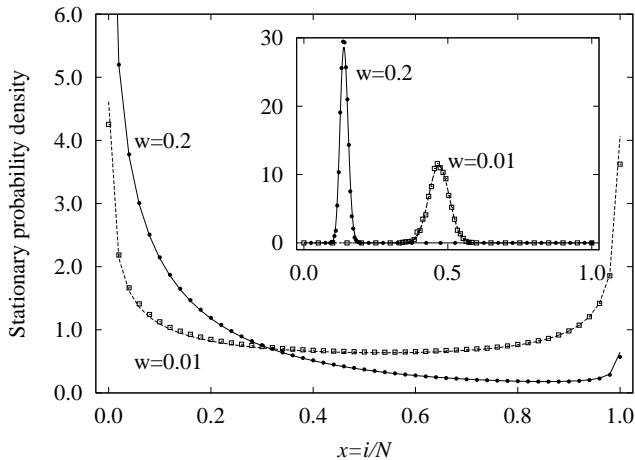}
\caption{
Stationary distribution of strategies in a Prisoner's Dilemma under the influence of mutations and selection in a finite population. The distribution undergoes a qualitative change with increasing population size. 
Main figure: 
For $N=50$, the stationary distribution for $w=0.2$ has a maximum at the Nash equilibrium $x=0$, where only defectors are present, as $u$ is below the critical mutation rate $u_c =0.02$. 
When selection is weak, it resembles the result from neutral selection ($w=0.01$). 
Inset: When the population size is increased to $N=10000$, the mutation rate is above the critical value $u_c \approx 10^{-4}$ and the situation resembles the result expected from the deterministic description of the replicator-mutator equation: The stationary distribution has a maximum at $x^{\ast} \approx 0.14$ for $w=0.2$, which is the stable fixed point of the replicator-mutator equation.  For weak selection, $w=0.01$, the stable fixed point is located at $x^{\ast} \approx 0.47$ and the distribution around this fixed point is wider due to stronger fluctuations. 
In all cases, numerical simulations of the frequency dependent Moran process (symbols) agree well with the analytical solution Eq.\ (\ref{statdist}) for the stationary distribution depicted by lines
($b=1$, $c=0.25$, mutation rate $u=0.01$, average over $10^8$ time steps).
}
\label{pd}
\end{figure}
The fitness is now a linear combination of the background fitness and the average payoff from interactions. For cooperators, the fitness becomes $\pi_C = 1-w+w ((b-c) (i-1) -c\,(N-i))/N$, whereas defectors have fitness $\pi_D = 1-w+w \, b \, i /N$. Since the fitness of defectors is always higher than the fitness of cooperators, $\pi_D>\pi_C $, any individual is better off not cooperating, despite the fact that mutual cooperation leads to a higher average payoff.  Without mutations the majority of stochastic trajectories would therefore end up in the absorbing state with $100 \%$ defectors and only few trajectories end up in pure cooperation. 
For $u <u_c=  1/(N+2)$, the stationary distribution keeps a maximum at $x=0$, but also a second (smaller) maximum at the other absorbing state $x=1$, see Fig.~\ref{pd}. For $u > u_c$, mutations drive the system away from these states and a maximum appears for intermediate values of $x$. The position of this maximum is determined by the intensity of selection $w$ and the payoff matrix.  For increasing $N$, the situation may change from $u<u_c$ to $u>u_c$. Therefore, the stochastic effects arising from the finite population size are clearly visible for small $N$, whereas large $N$ resembles a stationary distribution that is similar to the one expected from the deterministic replicator-mutator equation, see Sec.~\ref{repmuts}.

As a second example, we consider the Snowdrift game \cite{hauert:2004bo} also known as "Hawk-Dove game" or "Chicken". In this game, two players again choose between cooperation and defection. While the payoffs for defectors are as in the Prisoner's Dilemma, cooperation becomes more favorable: The payoff of a cooperator against a defector is now $b-c$ and therefore higher than the payoff of a defector against a defector. On the other hand, it is still lower than the payoff for mutual cooperation, $b-c/2$. The payoff matrix is given by 
\begin{equation}
M = 
\left(
\begin{array}{cc} 
m_{11} & m_{12}  \\
m_{21} & m_{22} 
\end{array}
\right)
=
\left(
\begin{array}{cc} 
b-\frac{c}{2}& b-c  \\
b & 0 
\end{array}
\right).
\end{equation}
\begin{figure}[thbp]
\includegraphics[totalheight=9cm,angle=270]{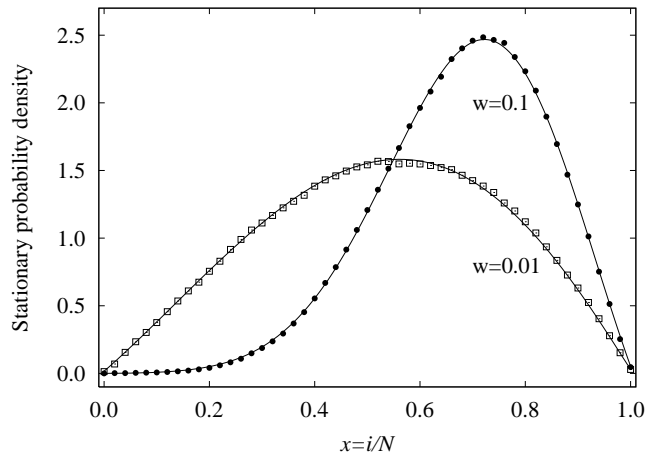}
\caption{
The Snowdrift game exhibits a stable interior equilibrium and typically coexistence of the two different types is observed. Only for small mutation rates and weak selection, the absorbing states are reached. Therefore, the stationary distribution has a maximum in the interior. For weak selection ($w=0.01$) this maximum is close to $x=0.5$. With stronger selection, the maximum moves toward the mixed equilibrium at $x^{\ast} \approx 0.856$. Numerical simulations (symbols) are in good agreement with the theoretical result depicted by full lines ($b=1$, $c=0.25$, mutation rate $u=0.01$, averages over $10^8$ time steps, $N=200$).}
\label{sd}
\end{figure}
Again, the payoff is a linear combination of the background fitness and the average payoff from interactions. 
In the Snowdrift game, rare strategies are always favored, resulting in a drift away from the absorbing states $x=0$ and $x=1$. 
Instead, the distribution is centered around a stable equilibrium in the interior. With decreasing intensity of selection, the distribution around this equilibrium becomes wider. 
In the limit $N \to \infty$ and without mutations ($u=0$), the distribution narrows and converges to $x^{\ast} = 2(b-c)/(2b-c)$, the only stable fixed point of the replicator dynamics. 
Finite size fluctuations lead to a distribution around this deterministic fixed point while mutations move the maximum of this distribution toward $x=0.5$. As for the Prisoner's Dilemma, simulations of the stationary distribution agree well with the analytical result Eq.\ (\ref{statdist}), cf.\ Fig. \ref{sd}.

\section{Replicator-Mutator equation}
\label{repmuts}
\vspace*{-1ex}
In the limit of infinite population size, our framework establishes a natural connection to the traditional deterministic description of coevolutionary dynamics. 
The deterministic nature of this limit arises from the fact that the diffusion matrix ${b_{kj}({\boldsymbol x})}$ and thus the finite size fluctuations vanish with $1/\sqrt{N}$ as $N \to \infty$. Hence, the Langevin equation becomes deterministic and the dynamics is given by the drift term, 
$\dot x_k =  \sum_{k=1}^d \left( T_{jk} -  T_{kj} \right)$. Since the mutation rate per
individual is fixed, the effect of mutations prevails in the limit of $N \to \infty$. 
In this case, Eq.\ (\ref{langevin}) incorporating mutation and selection can be written as
\begin{eqnarray}
\label{adrepmut}
\dot x_k 
& = & \frac{1}{\phi} \sum_{j=1}^d x_j \pi_j({\boldsymbol x}) q_{jk}-x_k.
\end{eqnarray} 
This is the replicator-mutator equation corresponding to the adjusted replicator dynamics \cite{maynard-smith:1982to}. 
So far, replicator-mutator equations \cite{bomze:1995rm,page:2002an} have not been derived from a microscopic birth-death process. They are the natural extension of the deterministic replicator equation \cite{taylor:1978wv,hofbauer:1998mm} that take into account mutations between strategies. Without mutations, $q_{jk}= \delta_{jk}$, types that have a fitness above the average fitness in the population increase in abundance, while types with a fitness below average decrease in abundance. Mutations interfere with this selection dynamics: Only types that have a fitness above the average {\em and} that have a sufficiently small mutation rate increase in frequency. Hence, the stable fixed points of the system  represent a balance between selection dynamics and mutation dynamics. 

\section{Local update process}
\label{lurs}
\vspace*{-1ex}
So far, we have only considered the frequency dependent Moran process. 
However, our framework can also be applied to other coevolutionary processes.
As an example, we generalize the local update process \cite{traulsen:2005hp} to 
an arbitrary number of strategies. In the local update process, 
an individual $1$ is chosen at random and compares its payoff to another randomly chosen individual $2$. With probability 
\begin{equation}
p = \frac{1}{2} + \frac{w}{2} \frac{\pi_2-\pi_1}{\Delta \pi}
\label{localtrans}
\end{equation}
$1$ adapts the strategy of $2$ (where the constant $\Delta \pi$ is the maximal possible payoff difference and $w$ denotes the intensity of selection) \cite{fn01}.
For $d$ different types, the transition probabilities are given by 
\begin{equation}
T_{kj}({\boldsymbol x}) =  {x_k x_j}\left(\frac{1}{2} + \frac{w}{2} \frac{\pi_j ({\boldsymbol x}) -\pi_k ({\boldsymbol x})}{\Delta \pi} \right).
\end{equation}
From the transition probabilities, we can calculate the drift term 
$a_{k}({\boldsymbol x}) = (w/\Delta \pi) \; x_k \left( \pi_k({\boldsymbol x})- \phi \right)$, 
which is independent of $N$.
The diffusion term is independent of the payoffs and simplifies to 
$b_{jk}({\boldsymbol x}) =x_j \left( \delta_{jk}- x_k \right)/N$.  
Thus, up to a constant rescaling of time, we recover the standard replicator dynamics for $d$ types in the limit $N \to \infty$, 
\begin{equation}
\dot x_k  =  x_k \left( \pi_k({\boldsymbol x})- \phi \right). 
\label{repmut}
\end{equation}
Up to a {\em dynamical} rescaling of time, Eq.\ (\ref{repmut}) is identical to Eq.\ (\ref{adrepmut}) without mutations. The natural extension of Eq.\ (\ref{repmut}) that includes mutations (while retaining the time scale) is the standard replicator-mutator equation \cite{page:2002an}, 
\begin{equation}
\dot x_k = \sum_{j=1}^d x_j \pi_j({\boldsymbol x}) q_{jk} -x_k \phi 
= \sum_{j,l=1}^d x_j \pi_j({\boldsymbol x}) (x_l q_{jk} - x_k q_{kl}).
\end{equation}
For general $q_{jk}$, it is not possible to define a microscopic process in which the transition probabilities only depend on payoff differences that leads to this differential equation. 
Only for the special case of vanishing mutations $q_{jk} = \delta_{jk}$, this is possible. 

A microscopic mechanism involving spontaneous mutations has been proposed in \cite{helbing:1996aa}. This yields the transition rates \cite{fn02}
\begin{equation}
T_{kj}({\boldsymbol x}) =  {x_k x_j}\left(\frac{1}{2} + \frac{w}{2} \frac{\pi_j ({\boldsymbol x}) -\pi_k ({\boldsymbol x})}{\Delta \pi} \right)  + x_k {q_{kj}}.
\label{lurtransprob}
\end{equation}
As shown in \cite{helbing:1996aa}, the limit $N \to \infty$ results in the differential equation
\begin{equation}
\dot x_k = \frac{w}{\Delta \pi} x_k \left(\pi_k({\boldsymbol x}) - \phi \right)
+ \sum_{j=1}^d \left( x_j q_{jk} - x_k q_{kj} \right).
\end{equation}
This selection mutation equation reduces to the standard replicator dynamics for $q_{jk} = \delta_{jk}$ \cite{hofbauer:1985jm,hofbauer:1998mm}. It describes selection under the influence of {\em spontanteous} mutations. Such spontaneous mutations can also be incorporated into other coevolutionary processes, leading to the same additive mutation term $\sum_{j=1}^d \left( x_j q_{jk} - x_k q_{kj} \right)$.

\section{Discussion}
\vspace*{-1ex}
Traditional descriptions of co-evolutionary systems rely on the assumption that populations are sufficiently large such that they can safely be considered infinite. In this case, accidental fixation of types with low fitness is not possible. For the more realistic case of finite populations, only few analytical methods exist that are often restricted to two types. Hence, these systems often rely on simulations which are very illuminative for special cases, but give little insight into the general dynamics of these systems. Here, we have shown how to extend the replicator dynamics in order to account for fluctuations arising from finite population sizes for any number of strategies and arbitrary mutation rates. When allowing for mutations, homogeneous states in which only one type is present are no longer absorbing boundaries and, instead, the system converges to a stationary distribution. This stationary distribution is derived using standard methods from statistical physics and depends on the population size, the mutation rate and the interaction parameters, i.e., the game. 
The fluctuation term allows for the calculation of corrections to the replicator dynamics for finite populations and is an important step in comparing individual based coevolutionary systems with deterministic dynamics. 
Mutations are not a source of fluctuations, but provide a mechanism that leads to a mixing of different types. However, fluctuations do not arise solely from finite populations. Instead, there are additional sources of noise that might have to be incorporated in other ways, such as noise from heterogeneity of agents or external noise that can influence interactions.

Our approach can be applied to any number of strategies $d$, but is based on the assumption that the population size is sufficiently large such that population densities can be approximated by smooth functions. In high dimensional spaces, such a continuum description may not always be appropriate \cite{tsimring:1996aa}.

Starting from microscopic descriptions for individual based coevolutionary processes with an arbitrary number of strategies, we have derived a replicator mutator equation with additional first order corrections that describe stochastic effects arising from finite populations.

\clearpage
\appendix
\section{Derivation of the Fokker-Planck equation for an arbitrary number of strategic types}
\vspace*{-1ex}
For $d$ strategies, the Master equation describing the change of the probability $P^{\tau}(i_1,..,i_d)$ that the system is in focal state $(i_1,..,i_d)$ is given by 
\begin{eqnarray}
\!\!\!\!
P^{\tau+\Delta \tau}(i_1,..,i_d) 
& \!\!=\!\! & P^{\tau}(i_1,...,i_d) \\ \nonumber
&& \! +  \Delta \tau \! \sum_{j,k=1}^d  P^{\tau}(i_1,..., i_j\!-\!1,...,i_k\!+\!1,...,i_d)  \\ \nonumber
&&  ~~~~~~~~~~~ \times   T_{kj}(i_1,..., i_j\!-\!1,...,i_k\!+\!1,...,i_d) \\ \nonumber
&& \! -  \Delta \tau \! \sum_{j,k=1}^d  P^{\tau}(i_1,...,i_d)  
T_{k  j}(i_1,...,i_d).
\end{eqnarray}
The first term is the probability that during a short time interval $\Delta \tau$ the state remains unchanged, the second term is the influx from other states and the third term is the outflux to other states. 
The probability $T_{k  j}(i_1,...,i_d)$ that one individual of type $k$ is replaced by another individual of type $j$ does not have to be specified. Special cases are given by Eqs.\ (\ref{Morantransprob}) and (\ref{lurtransprob}).
Next, we introduce the notation $x_l=i_l/N$ and $t=\tau/N$ and the probability density 
$\rho({\boldsymbol x};t)=N P^t({\boldsymbol x})$. For simplicity, we use the abbreviation
$(j^-,k^+)=(x_1,...x_j-{1}/{N},...,x_k+{1}/{N},...x_d)$ for states neighboring the focal state ${\boldsymbol x} = (i_1, \ldots , i_d) /N$ and obtain
\begin{eqnarray}
\frac{\rho({\boldsymbol x};t+\Delta t) - \rho({\boldsymbol x};t)}{N \Delta t} & = &  \sum_{j,k=1}^d  \rho(j^-,k^+;t)   T_{k  j}(j^-,k^+) \nonumber \\
&&- \sum_{j,k=1}^d  \rho({\boldsymbol x};t)  T_{k  j}({\boldsymbol x}),
\label{dFPG}
\end{eqnarray}
For $N \gg 1$, the left hand side can be approximated by $ \dot \rho({\boldsymbol x};t)/N$.
In order to expand the right hand side of Eq.\ (\ref{dFPG}) for $N \gg 1$, we rewrite the first term as
\begin{widetext}
\begin{eqnarray}
\sum_{j,k=1}^d  \rho(j^-,k^+;t)   T_{k  j}(j^-,k^+)   =  \sum_{j,k=1}^d 
\frac{ \rho(j^-,k^+;t)   T_{k  j}(j^-,k^+) +  \rho(j^+,k^-;t)   T_{j k}(j^+,k^-)}{2}.
\end{eqnarray}
The expansion of the probability density at a state $(j^{\pm}, k^{\mp})$ 
at time $t$ can be written as
 \begin{eqnarray}
\rho(j^{\pm},k^{\mp};t)  \approx  \rho({\boldsymbol x};t)
 \pm  \frac{1}{N} \left[\frac{\partial \rho({\boldsymbol x};t)}{\partial x_j} - \frac{\partial \rho({\boldsymbol x};t)}{\partial x_k}\right] 
+  \frac{1}{2 N^2} \left[\frac{\partial^2 \rho({\boldsymbol x};t)}{\partial x_j^2} -  2\frac{\partial^2 \rho({\boldsymbol x};t)}{\partial x_j \partial x_k}
 +\frac{\partial^2 \rho({\boldsymbol x};t)}{\partial x_k^2}\right] 
\end{eqnarray}
and similarly for the transition probabilities
 \begin{eqnarray}
T_{k j}(j^{\pm},k^{\mp})  \approx  T_{k j}({\boldsymbol x})
 \pm  \frac{1}{N} \left[\frac{\partial T_{k j}({\boldsymbol x})}{\partial x_j} - \frac{\partial T_{k j}({\boldsymbol x})}{\partial x_k}\right] 
+  \frac{1}{2 N^2} \left[\frac{\partial^2 T_{k j}({\boldsymbol x})}{\partial x_j^2} -  2\frac{\partial^2 T_{k j}({\boldsymbol x})}{\partial x_j \partial x_k}
 +\frac{\partial^2 T_{k j}({\boldsymbol x})}{\partial x_k^2}\right].
\end{eqnarray}
Let us consider the terms on the right hand side of Eq.\ (\ref{dFPG}) depending on their order in $1/N$. Obviously, the terms that are independent of $N$ cancel. Up to $1/N$, we find
\begin{eqnarray} 
&& \frac{1}{2N}\sum_{j,k=1}^d \Bigg[ \nonumber
\rho({\boldsymbol x};t) \left(\frac{\partial T_{k j}({\boldsymbol x})}{\partial x_k}-\frac{\partial T_{k j}({\boldsymbol x})}{\partial x_j} \right)
+ \rho({\boldsymbol x};t) \left(\frac{\partial T_{j k}({\boldsymbol x})}{\partial x_j}-\frac{\partial T_{j k}({\boldsymbol x})}{\partial x_k} \right)
+ T_{k j}({\boldsymbol x}) \left( \frac{ \partial \rho({\boldsymbol x};t)}{\partial x_k}-\frac{\partial \rho({\boldsymbol x};t)}{\partial x_j} \right)
\\ 
&&+ T_{j k}({\boldsymbol x}) \left(\frac{ \partial \rho({\boldsymbol x};t)}{\partial x_j}-\frac{\partial \rho({\boldsymbol x};t)}{\partial x_k} \right)
\Bigg] 
= -  \frac{1}{N}\sum_{k=1}^d \frac{\partial}{\partial x_k} \sum_{j=1}^d
\rho({\boldsymbol x};t) \left[  T_{j k}({\boldsymbol x})- T_{k j}({\boldsymbol x}) \right]
\label{driftcomp}
\end{eqnarray}
Since $1/N$ appearing on both sides of Eq.\ (\ref{dFPG}) cancels, this determines the drift term that is independent of $N$. All other terms vanish for $N \to \infty$. The first correction for finite
population sizes is the diffusion term resulting from terms in the expansion of the right hand side of Eq.~(\ref{dFPG}) that are of order $1/N^{2}$. For these terms, we find
\begin{eqnarray} 
&& \frac{1}{2N^2}\sum_{j,k=1}^d \Bigg[ \nonumber 
\frac{T_{k j}({\boldsymbol x}) + T_{j k}({\boldsymbol x})}{2} 
  \left(\frac{ \partial^2 }{\partial x_j^2} - 2 \frac{ \partial^2}{\partial x_j \partial x_k}
+\frac{ \partial^2}{\partial x_k^2}\right) \rho({\boldsymbol x};t) 
 +\rho({\boldsymbol x};t) \left( \frac{ \partial^2 }{\partial x_j^2} - 2 \frac{ \partial^2}{\partial x_j \partial x_k}
+\frac{ \partial^2}{\partial x_k^2}
\right)
\frac{T_{k j}({\boldsymbol x}) + T_{j k}({\boldsymbol x})}{2} \\ \nonumber
&& +\frac{\partial \rho({\boldsymbol x};t)}{\partial x_k} \frac{\partial T_{k j}({\boldsymbol x})}{\partial x_k}
-\frac{\partial \rho({\boldsymbol x};t)}{\partial x_k} \frac{\partial T_{k j}({\boldsymbol x})}{\partial x_j} 
-\frac{\partial \rho({\boldsymbol x};t)}{\partial x_j} \frac{\partial T_{k j}({\boldsymbol x})}{\partial x_k}
+\frac{\partial \rho({\boldsymbol x};t)}{\partial x_j} \frac{\partial T_{k j}({\boldsymbol x})}{\partial x_j}
 +\frac{\partial \rho({\boldsymbol x};t)}{\partial x_k} \frac{\partial T_{j k}({\boldsymbol x})}{\partial x_k}
-\frac{\partial \rho({\boldsymbol x};t)}{\partial x_k} \frac{\partial T_{j k}({\boldsymbol x})}{\partial x_j} \\ 
\label{diffcomp}
& & -\frac{\partial \rho({\boldsymbol x};t)}{\partial x_j} \frac{\partial T_{j k}({\boldsymbol x})}{\partial x_k}
+\frac{\partial \rho({\boldsymbol x};t)}{\partial x_j} \frac{\partial T_{j k}({\boldsymbol x})}{\partial x_j} \Bigg] =   \frac{1}{2N^2}\! \sum_{k,j=1}^d \! \! \Bigg[\! \!\left(\frac{\partial^2}{\partial x_k^2} - \frac{\partial^2}{\partial x_j \partial x_k} \right) 
\rho({\boldsymbol x};t) \left(
T_{k j}({\boldsymbol x}) + T_{j k}({\boldsymbol x})\right)   \Bigg].
\end{eqnarray}
\clearpage
\end{widetext}
So far, we have not used the normalization condition,  $\sum_{j=1}^{d} x_j=1$. Since $x_d$ is fully determined by $x_j$ with 
$1 \leq j < d$, we eliminate the dependence on $x_d$ in the transition probabilities $T_{jk}$ and in the probability density $\rho$. Hence, all derivatives with respect to $x_d$ vanish. Applying this and combining (\ref{driftcomp}) and (\ref{diffcomp}) with the expansion of the left hand side of Eq.\ (\ref{dFPG}), we obtain the Fokker-Planck equation
\begin{equation}
\dot \rho({\boldsymbol x})  = 
- \sum_{k=1}^{d-1} \frac{\partial}{\partial x_k} \rho({\boldsymbol x}) a_{k}({\boldsymbol x}) 
+ \frac{1}{2}\sum_{j,k=1}^{d-1} \frac{\partial^2}{\partial x_k \partial x_j}\rho({\boldsymbol x}) 
b_{jk}({\boldsymbol x})
\end{equation}
with drift vector
\begin{equation}
a_{k}({\boldsymbol x}) = \sum_{j=1}^{d} \left(T_{j k}({\boldsymbol x}) - T_{k j}({\boldsymbol x})\right)
\end{equation}
and a diffusion matrix $b_{jl}$ defined as
\begin{equation}
b_{jk}({\boldsymbol x}) \!\!
= \!\! \frac{1}{N} \!\!\left[ -T_{j k}({\boldsymbol x}) - T_{k j}({\boldsymbol x}) +\delta_{jk} \sum_{l=1}^d
T_{j l}({\boldsymbol x})+T_{l j}({\boldsymbol x}) \!
 \right],
\end{equation}
which is symmetric, $ b_{jk}({\boldsymbol x}) = b_{kj}({\boldsymbol x})$.
Note that the drift vector of a system with $d$ strategies has $d-1$ components, $k=1, \ldots, d-1$ and that the diffusion matrix is of dimension $d-1$. 
\vspace*{-6ex}
\acknowledgments{A.T.\ acknowledges support by the ``Deutsche Akademie der Naturforscher Leopoldina'' (Grant No.\ BMBF-LPD 9901/8-134).}
\vspace*{-1ex}
\bibliographystyle{h-physrev3}

\begin{thebibliography}{10}
\bibitem{fisher:1930}
R.~A.~Fisher,
{\em The Genetical Theory of Natural Selection}
(Clarendon Press, Oxford, 1930).

\bibitem{wright:1931}
S.~Wright,
Genetics
{\bf 16}, 97 (1931).

\bibitem{moran:1962ef}
P.~A.~P. Moran,
\newblock {\em The Statistical Processes of Evolutionary Theory} (Clarendon,
  Oxford, 1962).

\bibitem{kimura:1968aa}
M. Kimura, 
  { Nature} {\bf 217}, 624 (1968). 

\bibitem{peng:2004aa}
W. Peng, H. Levine, T. Hwa, and D. A. Kessler,
{Phys. Rev. E} {\bf 69}, 051911 (2004). 

\bibitem{cohen:2005aa}
E. Cohen, D. A. Kessler, and H. Levine,
{Phys. Rev. Lett.} {\bf 94}, 098102 (2005). 

\bibitem{maynard-smith:1982to}
J.~Maynard~Smith,
\newblock {\em Evolution and the Theory of Games} (Cambridge University Press,
  Cambridge, 1982).

\bibitem{nowak:2004aa}
M.~A. Nowak and K.~Sigmund,
\newblock Science {\bf 303}, 793 (2004).

\bibitem{taylor:1978wv}
P.~D. Taylor and L.~Jonker,
\newblock Math. Biosci. {\bf 40}, 145 (1978).

\bibitem{hofbauer:1998mm}
J.~Hofbauer and K.~Sigmund,
\newblock {\em Evolutionary Games and Population Dynamics} (Cambridge
  University Press, Cambridge, 1998).

\bibitem{bomze:1995rm}
I.~Bomze and R.~Buerger,
\newblock Games and Econ. Behav. {\bf 11}, 146 (1995).

\bibitem{page:2002an}
K.~M. Page and M.~A. Nowak,
\newblock J. Theo. Biol. {\bf 219}, 93 (2002).


\bibitem{nowak:2004pw}
M.~A. Nowak, A.~Sasaki, C.~Taylor, and D.~Fudenberg,
\newblock Nature {\bf 428}, 646 (2004).

\bibitem{taylor:2004wv}
C.~Taylor, D.~Fudenberg, A.~Sasaki, and M.~A. Nowak,
\newblock Bull. Math. Biol. {\bf 66}, 1621 (2004).

\bibitem{antal:2005aa}
T.~Antal and I.~Scheuring,
q-bio.PE/0509008 (2005).

\bibitem{imhof:2005oz}
L.~A. Imhof, D.~Fudenberg, and M.~A. Nowak,
\newblock Proc. Natl. Acad. Sci. USA {\bf 102}, 10797 (2005).

\bibitem{nowak:1992pw}
M.~A. Nowak and R.~M. May,
\newblock Nature {\bf 359}, 826 (1992).

\bibitem{lindgren:1994to}
K.~Lindgren and M.~G. Nordahl,
\newblock Physica D {\bf 75}, 292 (1994).

\bibitem{szabo:1998wv}
G.~Szab{\'{o}} and C.~T{\H{o}}ke,
\newblock Phys.\ Rev.\ E {\bf 58}, 69 (1998).

\bibitem{ebel:2002aa}
H.~Ebel and S.~Bornholdt,
\newblock Phys. Rev. E {\bf 66}, 056118 (2002).

\bibitem{szabo:2002mf}
G.~Szab{\'o} and C.~Hauert,
\newblock Phys. Rev. E {\bf 66}, 062903 (2002).

\bibitem{vukov:2005fa}
J.~Vukov and G.~Szab{\'o},
\newblock Phys. Rev. E {\bf 71}, 036133 (2005).

\bibitem{santos:2005bb}
F.~C. Santos, J.~F. Rodrigues, and J.~M. Pacheco,
\newblock Phys. Rev. E {\bf 72}, 056128 (2005).

\bibitem{santos:2006pn}
F.~C. Santos, J.~Pacheco, and T.~Lenaerts,
\newblock Proc. Natl. Acad. Sci. U.S.A. {\bf 103}, 3490 (2006).

\bibitem{ohtsuki:2006na}
H.~Ohtsuki, C.~Hauert, E.~Lieberman, and M.~A. Nowak,
\newblock Nature {\bf 441}, 502-505 (2006).

\bibitem{szabo:2002te}
G.~Szab{\'o} and C.~Hauert,
\newblock Phys. Rev. Lett. {\bf 89}, 118101 (2002).

\bibitem{lieberman:2005qx}
E.~Lieberman, C.~Hauert, and M.~A. Nowak,
\newblock Nature {\bf 433}, 312 (2005).

\bibitem{antal:2006pr}
T.~Antal, S.~Redner, and V.~Sood,
{Phys. Rev. Lett.} {\bf 96}, 188104 (2006).

\bibitem{traulsen:2005hp}
A.~Traulsen, J.~C. Claussen, and C.~Hauert,
\newblock Phys. Rev. Lett. {\bf 95}, 238701 (2005).

\bibitem{gardiner:1985bv}
C.~W. Gardiner,
\newblock {\em Handbook of Stochastic Methods}, 2nd ed. (Springer, Berlin,
  1985).

\bibitem{honerkamp:1994mm}
J.~Honerkamp,
\newblock {\em Stochastic Dynamical Systems} (VCH, Berlin, 1994).

\bibitem{kampen:1997xg}
N.~G.~v. Kampen,
\newblock {\em Stochastic Processes in Physics and Chemistry}, 2 ed. (Elsevier,
  Amsterdam, 1997).

\bibitem{imhof:2006ee}
L.~A. Imhof and D.~Fudenberg,
\newblock Journal of Economic Theory {\bf in press} (2006).

\bibitem{claussen:2005eh}
J.~C. Claussen and A.~Traulsen,
\newblock Phys. Rev. E {\bf 71}, 025101(R) (2005).

\bibitem{crow:1970ck}
J.~F. Crow and M.~Kimura,
\newblock {\em An Introduction to Population Genetics Theory} (Harper and Row,
  New York, NY, 1970).

\bibitem{rapoport:1965pd}
A.~Rapoport and A.~M. Chammah,
\newblock {\em Prisoner's Dilemma} (Univ. of Michigan Press, Ann Arbor, 1965).

\bibitem{axelrod:1984yo}
R.~Axelrod,
\newblock {\em The Evolution of Cooperation} (Basic Books, New York, 1984).

\bibitem{hauert:2004bo}
C.~Hauert and M.~Doebeli,
\newblock Nature {\bf 428}, 643 (2004).

\bibitem{fn01}
{If additionally $a$ adopts strategy $b$ with probability $1-p$, the dynamics becomes twice as fast and fluctuations (i.e.\ the term $c_{kj}({\boldsymbol x})$ in Eq.~(\ref{langevin})) increase by a factor $\sqrt{2}$. }

\bibitem{helbing:1996aa}
D.~Helbing,
\newblock Theory and Decision {\bf 40}, 149 (1996).

\bibitem{fn02}
The transition rates $T_{kj}$ can be normalized to probabilities, 
changing only the timescale in the limit $N \to \infty$.

\bibitem{hofbauer:1985jm}
J.~Hofbauer,
\newblock Journal of Mathematical Biology {\bf 23}, 41 (1985).

\bibitem{tsimring:1996aa}
L.S. Tsimring, H. Levine, and D.A. Kessler,
Phys. Rev. Lett. {\bf 76}, 4440 (1996). 

\end{thebibliography}

\end{document}